\documentstyle[12pt]{article}

\catcode`@=11
\def\relstack#1#2{\mathrel{\mathop{#1}\limits_{#2}}}
\catcode`@=12

\begin{document}
%
%
\newcommand{\ECM}{\em Dept. Estructura i Constituents de la Mat\`eria
                  \\ \em Fac. de F\'\i sica, Universitat de Barcelona \\
                     \em  Diagonal 647, 08028 Barcelona (Spain) \\ and
\\ I.F.A.E.}

%
%
\newcommand{\NPB}[3]{{\em Nucl. Phys.} {\bf B#1}, {#2} (19#3)}
\newcommand{\PRD}[3]{{\em Phys. Rev.} {\bf D#1}, {#2} (19#3)}
\newcommand{\PLB}[3]{{\em Phys. Lett.} {\bf B#1}, {#2} (19#3)}
\newcommand{\PRL}[3]{{\em Phys. Rev. Lett.} {\bf #1}, {#2} (19#3)}
\newcommand{\AJP}[3]{{\em Am. J. Phys.} {\bf #1}, {#2}
(19#3)}
\newcommand{\PR}[3]{{\em Phys. Rev.} {\bf #1}, {#2} (19#3)}
\newcommand{\AP}[3]{{\em Ann. of Phys. (N.Y.)} {\bf #1}, {#2} (19#3)}
\newcommand{\IJA}[3]{{\em Int. Journ. of Mod. Phys.} {\bf A#1}, {#2}
(19#3)}

\newcommand{\CMP}[3]{{\em Comm. Math. Phys.} {\bf #1}, {#2} (19#3)}
\newcommand{\beq}{\begin{equation}}
\newcommand{\eeq}{\end{equation}}

\pagestyle{empty}
\vspace*{-2cm}
{\hfill \parbox{6cm}{\begin{center} UB-ECM-PF 19/93\\
                                    September 1993
                     \end{center}}}
\vspace{1.5cm}

\begin{center}
\large{\bf Perturbative Renormalization in Quantum Mechanics}
\end{center}
\vskip .6truein
\centerline {Cristina Manuel \footnote[1]{Bitnet PALAS@EBUBECM1}
             and Rolf Tarrach \footnote[2]{Bitnet ROLF@EBUBECM1}}
\begin{center}
 \ECM
\end{center}
\vspace{1.5cm}

\centerline{\bf Abstract}
\medskip
Some quantum mechanical potentials, singular at short distances, lead to
ultraviolet divergences when used in perturbation theory. Exactly as in
quantum field theory, but much simpler, regularization and
renormalization lead to finite physical results, which compare correctly
to the exact ones. The Dirac delta potential, because of its relevance
to
triviality, and the Aharonov-Bohm potential, because of its relevance to
anyons, are used as examples here.


\newpage
\pagestyle{plain}

An important instrument of present days physics, quantum field theory,
is permeated by short distance singularities, which are thoroughly
understood in the framework of regularized and renormalized perturbation
theory. Our non-perturbative understanding, mainly lattice-bound, is not
so firm, and exact solutions for physically relevant theories are
basically absent. Quantum mechanics does not usually have short distance
singularities, but they show up if the potential is singular enough (but
not too much: the Hamiltonian should be bounded from below and
self-adjoint). Then regularization and renormalization consistently cure
the short distance singularities and lead to physical results
independent of the precise regulator and independent of the precise
renormalization scheme. Furthermore, renormalized perturbation theory
reproduces the exact solutions for physical magnitudes.

The two problems we have chosen to study are best considered in two
dimensions. First, the Dirac delta, zero-range or contact interaction,
because already its exact solution is most conveniently obtained by
regulating and renormalizing, because one can perform perturbation
theory to all orders, and because of its relevance to triviality
\cite{Calla}.
Second, the Aharonov-Bohm potential, because it perturbatively induces a
new interaction absent in the exact setting and because of being at the
foundation of anyon physics \cite{Lein}, \cite{Wilczek}.

Recall that the Schr\"{o}dinger equation (throughout this
letter we will use $2M=\hbar=1$)
for positive energies $E=k^2$ is equivalent to
 the Lippman-Schwinger equation
$$\psi(\vec{r}) = \psi_0 (\vec{r}) - \int d^2 r' G_{k,+}
(\vec{r}-\vec{r'}) V(\vec{r'}) \psi (\vec{r'}),$$
where $\psi_0$ is a solution of the free Schr\"{o}dinger equation, and
$G_{k,+}$ is the free propagator.
In the two dimensional case that we are interested in
$$
G_{k,+} (\vec{r} - \vec{r}^{'}) = \frac {i}{4} H_0 ^{(1)} \left (k
 |\vec{r}-\vec{r}^{'}| \right),
$$
$H_0 ^{(1)}$ being  the first Hankel function of zero order, which
in the asymptotic limit  behaves as an outgoing
wave and has the short distance behavior
\beq
G_{k,+} (\vec{\epsilon}) \relstack{\sim}{\epsilon \rightarrow 0}
\frac{i}{4}-
\frac{1}{2\pi} \left( \ln(\frac{k\epsilon}{2}) + \gamma \right),
\qquad
 \vec{\epsilon} \equiv \vec{r} - \vec{r'},
 \label{short.prop}
\eeq
where $\gamma$ is Euler's constant.
One particular way to find the solution of the Schr\"{o}dinger
equation consists
in solving iteratively the Lippman-Schwinger equation, so that
\begin{eqnarray}
\psi(\vec{r}) & = &\psi_0 (\vec{r}) - \int d^2 \vec{r'} G_{k,+} (\vec{r}
- \vec{r'}) V(\vec{r'}) \psi_0 (\vec{r'}) \nonumber \\
 & + & \int d^2 \vec{r'} d^2 \vec{r''} G_{k,+} (\vec{r}-\vec{r'})
V(\vec{r'}) G_{k,+}(\vec{r'}-\vec{r''})V(\vec{r''}) \psi_0 (\vec{r''})
+
...
\label{iterLip}
\end{eqnarray}
In the scattering problem, the above approach together with the
asymptotic condition
\beq
\psi(\vec{r}) \relstack {\sim}{r \rightarrow \infty} e^ {i \vec{k}
\cdot
\vec{r}} + \frac {e^ {i \left(kr + \frac{\pi}{4} \right)}}{\sqrt{r}}
f(k,\theta),
\label{asym.con.}
\eeq
allows to define the Born series to compute the scattering amplitude
$f$ in perturbation theory. In this context, it is easy to understand
that for a potential $V(\vec{r})$ which is singular enough
when $r \rightarrow 0$, the Born series will contain divergent
coefficients, as the propagator exhibits the short distance
logarithmic
behavior (\ref{short.prop}). We will illustrate this fact by studying
the perturbative approach of two different potentials which can be
solved exactly but have that peculiarity.

Let us focus first our attention on the perturbative approach to the
Dirac
delta potential, zero-range or contact interaction.
Let us recall that contact interactions
have been studied exactly in the literature \cite{Thorn}-\cite{crha}.
 Delta function
potentials in two and three dimensions are a nice and
simple example where the concepts of regularization and
renormalization, which are common in quantum field theory but hardly
used in quantum mechanics, are very useful for obtaining non-trivial
results. One correct treatment of the problem requires first
regularization. This can be achieved by substituting the delta function
potential $g_0 \delta^{(2)} (\vec{r})$, $g_0 <0$,
 into the Schr\"{o}dinger
equation by, say,
\beq
V_R (r) =   \frac{g_0}{2\pi R} \delta(r-R); \qquad R>0,
\label{displa}
\eeq
 which in
the $R \rightarrow 0$ limit  reproduces the original potential. The
Schr\"{o}dinger equation for negative energies
can then easily  be solved by well-known methods, finding for small
$R$ one binding
energy  $E_0 (R) = -\frac{4}{R^2} e^{-2 \gamma} e^{4 \pi / g_0}$.
For $R \rightarrow 0$ the bound state energy goes to $-\infty$.
 However, one can
choose a coupling constant depending on the regulator in such a way that
when the regulator is removed the binding energy stays finite. This
happens when
$$
\frac{1}{g_0 (R)} = \frac{1}{2 \pi} \ln{\frac{R}{R_0}}, \qquad R<R_0,
$$
where $R_0$ is a length which measures the strength of the interaction.
Because one is used to describe interactions with dimensionless
couplings one introduces the renormalized coupling constant $g_r(\mu)$,
which depends on an arbitrary momentum scale $\mu$,
\beq
\frac{1}{g_r (\mu)} \equiv \frac{1}{g_0 (R)} -
\frac{1}{2 \pi} \left(\ln{\frac{\mu R}{2}} + \gamma \right),
\label{ren.copl}
\eeq
in terms of which the  binding energy reads
$ E_0 = -\mu^2 e^{4 \pi / g_r(\mu)}$. Notice that $E_0$ depends on
$g_r(\mu)$, but not on $\mu$, that is, the explicit and implicit
(through $g_r$) dependence on $\mu$ of $E_0$ cancel. Physics is
determined by the value of $g_r(\mu)$ at an arbitrary value of $\mu$.

Similar features are found in the scattering problem, $E=k^2 >0$. Only
when the coupling constant depends on $R$ as above  is it
possible to find the finite and renormalized scattering amplitude
\beq
f(k, \theta)  = \left(\frac{\pi}{2k}\right)^{1/2} \frac
{1}{\ln{\frac{k}{\mu}}- \frac{i\pi}{2}-\frac{2\pi}{g_r(\mu)}}.
\label{exactres}
\eeq

Contact interactions have also been studied in the literature with
the use of some other regularizations, such as a circular well
potential \cite{Gosd}, \cite{Perez},
or in  momentum space \cite{Thorn}, \cite{Beg}, \cite{Jackiw},
 arriving to the conclusion
that  regularization and renormalization is a powerful approach for
obtaining non-trivial results which are furthermore independent of the
particular regulator used. It is also important to stress that there is
a large amount of arbitrariness in defining $g_r(\mu)$: a specific
definition, as (\ref{ren.copl}), defines a renormalization scheme, but
one could have added a constant to the r.h.s. of (\ref{ren.copl}) which
would modify the dependence of $E_0$ and of $f(k,\theta)$ in $g_r(\mu)$
without actually changing physics. This is seen immediately recalling
that the dependence of $f(k,\theta)$ on $g_r(\mu)$  and $\mu$ can both
be traded for a dependence on $E_0$, which is fixed by a renormalization
condition \cite{Gosd}.

{}From references \cite{Thorn}-\cite{crha}   we learn that the delta
function potential in
two space dimensions: i) is most appealingly treated by regularization
and renormalization of the coupling constant;  ii)
 exhibits the typical quantum scale anomaly associated to the process
of renormalization, since the problem was originally scale invariant
but the final answer is not; iii) is
only noticed by s-waves, since for higher waves the centrifugal barrier
dominates over the delta function potential, which therefore  is
 completely
screened; iv) can be substituted by a boundary condition
which ensures the self-adjoint character of the  radial free
Hamiltonian;
v) is the formal non-relativistic limits of a $g \phi^4$ quantum
field theory.

After recalling the basic features of contact interactions, we will
address here its perturbative approximation. One expects that
perturbative renormalization should be used to deal with this problem,
and in fact this is  so. If one substitutes
directly $g_0 \delta^{(2)} (\vec{r})$ in (\ref{iterLip}), the second
and higher order results are clearly logarithmically divergent. As we
already mentioned, this is due to the short distance behavior of the
propagator (\ref{short.prop}). To compute explicitly the divergences, we
will introduce the regulated expression for the delta function
potential given in (\ref{displa}).
The first term in the Born series of the scattering amplitude, if
$\vec{k}_i$ and $\vec{k}_f$ are the momenta of the incident and
scattered particles respectively, is
$$
f^{(1)}_B = - \frac {1}{2 \sqrt{2 \pi k}}
\int d^2 r e^{-i \vec{k}_f \cdot \vec{r}} V_R (\vec{r}) e^{i
\vec{k}_i \cdot \vec{r}}=
\frac {-g_P}{2 \sqrt{2
\pi k}}  + O(R),
$$
where we call the coupling in which we perturb $g_P$.
Although the first order approximation is finite when the regulator
is removed, the second order one is logarithmically divergent
\begin{eqnarray}
f^{(2)}_B & = &\frac{1}{2 \sqrt{2\pi k}} \int d^2 r' d^2 r'' e^{-i
\vec{k}_f \cdot \vec{r'}} V_R (\vec{r'}) G_{k,+}
(\vec{r'}-\vec{r''}) V_R (\vec{r''}) e^{i \vec{k}_i \cdot
\vec{r''}}  \nonumber \\
& = & \frac{1}{2 \sqrt{2 \pi k}} g^2 _P \left(\frac{i}{4} -\frac{1}{2
\pi} \ln{\frac {kRe^{\gamma}}{2}}   \right) + O(R).
\label{secondorder}
\end{eqnarray}
In order to get rid of this divergence in the scattering amplitude we
will choose another potential  given by
\beq
V_R (r)=V^{(1)}_R (r) + V^{(2)}_R (r) =  \frac{g_P}{2\pi R} \left( 1 -
\frac{g_P}{2\pi} \ln{\frac{R \Lambda e^{\gamma}}{2}} \right)
\delta(r-R), \label{twowell}
\eeq
where $\Lambda$ is an arbitrary momentum. The computation up to second
order in $g_P$ using the above potential leads to a finite answer,
since the second order in $g_P$ in (\ref{twowell}) exactly cancels the
logarithmic divergence in (\ref{secondorder}), and then
$$
f_B = \frac{-g_P}{2\sqrt{2\pi k}}  \left( 1 - g_P \left(\frac{i}{4}
-\frac{1}{2\pi}\ln{\frac{k }{\Lambda}}\right) \right) + O(g_P
^3).
$$
Notice that the modification introduced in (\ref{twowell}) is only aimed
at making the physical magnitudes finite when $R \rightarrow 0$, not at
introducing a new scale $\Lambda$. Thus $g_P$ has to depend on $\Lambda$
in such a way as to cancel the $\Lambda$ dependence of $f_B$.

 Perturbation theory up to third order in $g_P$ using
potential (\ref{twowell}) would yield  divergent quantities once again
$$
f_B = \frac{-g_P}{2\sqrt{2\pi k}} \left(1- g_P
\left(\frac{i}{4}-\frac{1}{2 \pi} \ln{\frac{k }{ \Lambda}} \right)
+ g_P ^2 \left(\frac{i}{4}- \frac{1}{2\pi} \ln{\frac{kRe^\gamma}{2}}
\right)^2 \right) + O(g_P^4).$$

The potential that leads to finite results up to
$g_P^3$ is given by
$$
V_R (r) =V_R ^{(1)}+ V_R^{(2)}+V_R ^{(3)}=  \frac{g_P}{2\pi R} \left( 1
- \frac{g_P}{2\pi} \ln{\frac{R \Lambda
e^\gamma}{2}} + \frac{g_P ^2}{4\pi^2} \ln^2 {\frac{R \Lambda
e^\gamma}{2}} \right) \delta(r-R),
$$
so that one obtains
$$
f_B = \frac{-g_P}{2\sqrt{2\pi k}} \left(1- g_P
\left(\frac{i}{4}-\frac{1}{2 \pi} \ln{ \frac{k }{ \Lambda}} \right) +
g_P ^2 \left(\frac{i}{4}- \frac{1}{2\pi} \ln{\frac{k}{\Lambda}}
\right)^2 \right)  + O(g_P ^4).
$$
One can easily compute to all orders in perturbation theory, and
the series obtained can be summed up giving
$$
V_R (r)  =   \frac{1}{2\pi R} \frac{g_P}{1 + \frac{g_P}{2 \pi}
\left(
\ln{\frac{R \Lambda}{2}}+ \gamma \right)} \delta(r-R).
$$
and
\beq
f_B = \left(\frac{\pi}{2k}\right)^{1/2} \frac {-g_P/2\pi}{1+g_P
\left(\frac{i}{4}-\frac{1}{2\pi} \ln{\frac{k}{\Lambda}} \right)}.
\eeq
The above result agrees with the exact one (\ref{exactres}) after
imposing $\Lambda e^{2 \pi /g_P} = \mu e^{2 \pi /g_r}$, which
implies $g_P(\Lambda = \mu) \equiv g_r (\mu)$. The bare coupling
constant is defined from $V_R(r)$ as
$$
g_0 (R) \equiv     \frac{g_P(\Lambda)}{1 + \frac{g_P(\Lambda)}{2 \pi}
\left( \ln{\frac{R \Lambda}{2}}+ \gamma \right)},
$$
which coincides with (\ref{ren.copl}).
This identification should be understood as a renormalization
condition, that in principle, one should impose at each order of the
perturbative expansion.

The use of the specific regulator we chose allows to easily
compute to all
orders the Born scattering amplitude and sum the perturbative results.
In general, with the use of some
other regulator (e.g. a  circular well potential), renormalized
perturbation theory does not reproduce both the exact regularized
potential and the exact scattering amplitude. This latter is easily
reproduced, which is all what physics demands, but the summed
regularized potential is not the same as the exact one.

This  analysis explains how the perturbative approach of contact
interactions in quantum mechanics has to be performed, which is close
both in spirit and technically to  standard perturbative
regularization and renormalization of quantum field theory.
 We have seen that in order to obtain finite
results up to a certain order in perturbation theory one has to add some
"counterterms" to the potential which cancel the logarithmic divergences
produced in the coefficients of the Born series.
Then these divergences are reabsorbed into the bare parameters of
the theory, while physical finite magnitudes can be defined.
This is exactly the
same process that is followed when renormalizing a quantum field theory
perturbatively.

We will address now the perturbative approach to the Aharonov-Bohm
problem, that also requires the techniques of renormalization.
The Aharonov-Bohm gauge potential is
$\vec{A} =  \frac {\Phi}  {2\pi r} \hat{ u}_{\varphi}$
where $\Phi$ is the magnetic flux, and $\hat{ u}_\varphi$ is the unitary
vector in the azimuthal direction and $r$ is the radial distance.
 We define $\alpha \equiv
-\frac{e\Phi}{2\pi}$, $e$ being the electric charge of the particle.
Restricting the study of the problem to $0<\alpha<1$, the exact
scattering amplitude of this problem has been computed by several
authors \cite{Bohm}, \cite{Ruij}, \cite{Hagen},  \cite{Jack}
 and for an incident wave with momentum $\vec{k}_i =(-k,0)$ is,
in the non-forward direction \footnote{ As noticed in the literature
\cite{Ruij}, \cite{Hagen},  \cite{Jack}, if the exact Aharonov-Bohm
scattering  wavefunction obeys the asymptotic condition
(\ref{asym.con.}), then the scattering amplitude exhibits a singular
contribution in the forward direction. This singularity can be removed
by modifying that asymptotic condition as it was done in \cite{Bohm}.
Here we will restrict our study to the non-foward direction, noticing
that also the perturbative approach yields singularities in the forward
direction.},
 $$
f(k, \theta) =- \frac{\sin \pi \alpha}{\sqrt{2 \pi k}} \left(1 - i\tan
\frac{\theta}{2} \right); \qquad     -\pi < \theta < \pi.
$$
The limit $\alpha \rightarrow 0$ can be taken in the above expression,
obtaining
\beq
f(k, \theta) =- \frac{\pi \alpha}{\sqrt{2 \pi k}} \left(1 - i\tan
\frac{\theta}{2} \right) + O(\alpha^3).
\label{exactAB}
\eeq
Notice that there is no $\alpha^2$ contribution in (\ref{exactAB}). This
is the result one would expect to recover using perturbation theory up
to second order. As remarked by several authors \cite{Cori},
\cite{Ahar}, \cite{Nagel},
 the
Born series fails to give the correct results. We will show here how
one should proceed to obtain the correct answers.

The perturbation  in this problem is
$$
V(\vec{r}) = -2 i \frac{\alpha}{r^2} \frac{\partial}{\partial \varphi}
+ \frac{\alpha^2}{r^2} \equiv V_{\alpha}^{(1)} (\vec{r}) +
V_{\alpha^2}^{(2)} (r). $$
The Born approximation in first order in $\alpha$ gives
$$
f_B^{(1)} = \frac{i \pi \alpha}{\sqrt{2 \pi k}} \tan \frac{\theta}{2}.
$$
This does not agree with the result obtained in (\ref{exactAB})
in first order in $\alpha$
since there is one term which is missing. As noticed by Corinaldesi and
Rafeli \cite{Cori}, in
the first order approximation the s-wave contribution to the scattering
amplitude is absent.

The correction in $\alpha^2$ comes from two terms. The first one, from
the iteration of $V_{\alpha}^{(1)}$, is finite, and can be computed best
by going to momentum space and computing the
integral
$$
f_B ^{(2),1} = \frac{4 \alpha^2}{2 \sqrt{2 \pi k}} \int \frac{p dp
d\varphi}{p^2 -k^2 -i \epsilon}
\frac{kp \sin \varphi}{(k^2+p^2 + 2kp\cos \varphi)}
\frac{kp \sin(\theta-\varphi)}{(k^2+p^2 - 2kp\cos(\varphi-\theta))}.
$$
The angular integral can be trivially evaluated by expanding each
fraction
in Gegenbauer polynomials. After some algebra, one arrives at
$$
f_B ^{(2),1} = -\frac{2 \pi \alpha^2}{2 \sqrt{2 \pi k}} \left( \ln(2
\cos \theta/2) + \frac{i \pi}{2} \right).
$$
The second contribution of order $\alpha^2$ to the Born scattering
amplitude comes from $V_{\alpha^2}^{(2)}$ and it yields a
logarithmically divergent integral
$$
f_B ^{(2),2} =-\frac{\alpha^2}{2\sqrt{2 \pi k}} \int d^2 r
\frac{e^{i(\vec{k}_i - \vec{k}_f) \cdot \vec{r}}}{r^2}
= -\frac{2\pi \alpha^2}{2\sqrt{2\pi k}} \int^{\infty} _{0} dr
\frac{ J_0(\kappa r)}{r},
$$
where $J_0$ is the Bessel function of zero order,
and $\kappa = |\vec{k}_i-\vec{k}_f
|= 2 k \cos \theta/2$. Introducing a short distance cut-off, we can
compute the explicit form of the divergence
$$
f_B ^{(2),2}
= -\frac{2\pi \alpha^2}{2\sqrt{2\pi k}} \left(-\gamma - \ln{\frac{\kappa
R}{2}} + O(R) \right).
$$
Adding all contributions up to second order in $\alpha^2$, the Born
scattering amplitude is
\beq
f_B = \frac{ \pi \alpha}{\sqrt{2 \pi k}} \left(i \tan \frac{\theta}{2}
+ \alpha \left( -i \frac{\pi}{2} + \gamma + \ln {\frac{kR}{2}}\right)
\right) + O(\alpha^3).
\label{2BornAB}
\eeq
This does not reproduce the correct
result in first order in $\alpha$ and yields a divergent quantity
in second order, even though the correct result does not have
contribution in $\alpha^2$. Notice, as well, the kind of divergence
in (\ref{2BornAB}): it is exactly of the same form as the one we found
in the second order perturbation of the contact
interaction, eq. (\ref{secondorder}).
In fact, introducing a new interaction in the perturbed
Hamiltonian, a zero-range or contact interaction $V(\vec{r})= 2\pi
\alpha \delta^{(2)}(\vec{r})$, the wrong first order result is
corrected, while the whole second order result, including
the logarithmic divergence, is cancelled, so that one
arrives at precisely the scattering amplitude (\ref{exactAB}).

We should remark here that the crossed terms in the scattering amplitude
in $\alpha^2$ corresponding to the  the delta function
potential times $V^{(1)}_{\alpha}$ trivially vanish,  the reason being
that they act in orthogonal Hilbert subspaces: while the contact
interaction only affects the s-wave sector, $V^{(1)}_{\alpha}$ vanishes
in that subspace.

The contact interaction introduced here is a repulsive one. This very
likely reflects its origin as an auxiliary interaction only seen in
perturbation theory, but which is not present in the exact treatment
(recall that repulsive contact interactions are trivial). It contributes
to implement a feature of the exact wavefunction, i.e. that it vanishes
at the origin, which perturbations in the Aharonov-Bohm potential alone
are not able to implement, leading to the short distance divergence of
$f_B ^{(2),2}$. Notice also that
the coupling constant of this interaction does not become
renormalized, since its unique job is to make finite the perturbative
theory of the Aharonov-Bohm problem. Obviously, one could add a
different contact interaction to the Hamiltonian and proceed with  its
perturbative study. This was studied exactly in ref. \cite{Man}.

We should mention here that similar results were first found studying
the non relativistic quantum field theory model corresponding to the
Aharonov-Bohm effect \cite{Ha}. In refs. \cite{Bergman}, \cite{vascos},
\cite{Freedman}
it was shown that in order that
the theory is finite a $\phi^4$ interaction is needed. We find parallel
results in quantum mechanics: the introduction of a new interaction in
the perturbed Hamiltonian, a delta function potential, which is the
quantum mechanical counterpart of $\phi^4$, is needed to ensure the
finiteness of the perturbative approach.

These results should also be applied to the study of the perturbative
theory of anyonic quantum mechanics in the bosonic end, where one
also finds logarithmic divergent quantities. This problem has been
considered recently by several authors \cite{McCabe}, \cite{Sen},
\cite{Chou},
 but we think that the solution
we offer here, not unrelated to some of the ones suggested by these
authors, is simple, systematic and straightforward.

In conclusion: quantum mechanics for singular enough potentials
leads to a perturbation theory anchored on regularization and
renormalization. It shows in a very simple setting many features of
renormalized perturbation theory in quantum field theory, but allows to
go far beyond by, i.e. comparing to the known exact results.

\vbox{}
\vfill
\hbox{}
{\large \bf Acknowledgements}
We thank X. Vilasis-Cardona, J.I. Latorre and R. Emparan for useful
discussions. Financial support under contract AEN90-0033 is
acknowledged. C. Manuel acknowledges the Ministerio de Educaci\'{o}n
y Ciencia for an FPI grant.

\end{document}